# A Novel Algorithm In Steganography Using Weighted Matching Technique

P N Priya[1], R Ranjitha[2], Yashaswini Naik[3], Shrilekha[4], Rama Moorthy H[5]

[1,2,3,4] Computer Science Engineering, SMVITM, Bantakal, Udupi-574115

[5] Assistant Professor, CSE, SMVITM, Bantakal, Udupi-574115

**Abstract—** *In recent years, the security related to data over the internet has become a major issue. Different techniques are used to secure the information, one such technique is steganography. In steganography, one can have a cover media as an image, video or audio. Here we are proposing a method of weighted matching technique in which we examine higher bits of lower nibble of a byte and find the highest occurrence bit. The PSNR values of both the algorithms are tabulated for evaluation purpose.*

## I. INTRODUCTION

Ensuring the data security is a big challenge for computer users. Businessmen, Professionals, Military and home users all have some important data that they want to secure from attacks [3]. There are number of ways for securing data. Steganography is one among them. Cryptography and steganography are often used together even though they are very unalike, yet complementary, technologies with different purpose. Cryptography is technique to scramble confidential information to make it "unreadable." It is already commonly used in Internet communications. Encrypting a message can hide content of the secret message effectively, but it cannot hide its presence. The location of secret message is very obvious hence it is the reason why an encrypted message can easily be targeted by attackers.

Steganography techniques are used to deal with digital copyrights managements, protect information, and conceal secret data. Data hiding techniques provide a motivating challenge for digital forensic investigators. The backbone of today's communication is data. To ensure that data is secured and does not go to undesired destination, the concept of data hiding came up to protect a part of information. Digital data can be delivered over computer networks with small amount of errors and often without interference. The Internet provides a communication method to distribute information to the masses. Hence, the privacy and data reliability are required to protect against unauthorized access and use. Steganography relies on hiding message in unsuspected multimedia data and generally used in secret communication between recognized parties. Here, the technique replaces unused or insignificant bits of the digital media with the secret data. The concept is to embed the hidden object into a considerably larger object so that the changes are undetectable by the human eye. All digital file formats can be used for steganography, but formats those are with a high scale of redundancy are more suitable. Audio and video steganography is least suitable when the secret message length is small hence image steganography is preferred.

## II. LITERATURE REVIEW

There are various kinds of techniques to achieve steganography like Least Significant Bit Insertion, Masking & Filtering and Algorithms & Transformations [4]. Each of these techniques can be applied, with varying degrees of success, to different images. Least Significant Bit Insertion is a simple approach to embedding information in a cover file. Section A deals with representation of images. The basics of LSB embedding and the analysis based on LSB embedding and algorithms and implementation are explained in section B**.** Drawback of LSB insertion is explained in section C.

### A. Representation of image types:

In a computer, images are represented as array of numbers that represent intensities of the three colors R(Red), G(Green) and B(Blue), where a value for each of the three colors describes a pixel. Through varying the intensities of the RGB values, a finite set of colors spanning the full visible spectrum can be created. In an 8-bit image, there can be $2^8 = 256$ colors and in a 24-bit bitmap, there can be $2^{24} = 1$ 6.7mill ion colors. Large images are most desirable for steganography because they have the most space to hide data in .The best quality hidden message is normally produced using a 24-bit bitmap as a cover image.

A 24-bit bitmap is bitmap header,

**1001010<u>1</u>  0000110<u>1</u>  1100100<u>1</u>  1001011<u>0</u>**

**0000111<u>1</u>  1100101<u>1</u>  1001111<u>1</u>  0001000<u>0</u>**

followed by the pixel's data. The higher the number, the more intense that color is for the pixel. The drawback of large images is that they are heavy for both transfer and upload, hence larger chance of drawing an "attacker's" attention due to their uncommon size. Hence, compression is often used [4].

*B. Basic of LSB Embedding:*

The concept of LSB Embedding is, it exploits the fact that the level of precision in many image formats is far greater than that perceivable by average human perception. Therefore, an altered image with slight variations in its colors will be indistinguishable from the original for humans, just by looking at it. As a simple example of least significant bit substitution, imagine "hiding"' the character **'F'** across the following eight bytes of a carrier image (the least significant bits are underlined): **'F'** is represented as the binary string **01000110** in **ASCII.** These eight bits can be written" to the least significant bit of each of eight carrier bytes as follows:

**1001010<u>0</u>   0000110<u>1</u>   1100100<u>0</u>   1001011<u>0</u>**

**0000111<u>0</u>   1100101<u>1</u>   1001111<u>1</u>   0001000<u>0</u>**

*C. Drawback LSB insertion method*

- Although LSB embedding method hide the data in such a way that human eye don't perceive it, it can be easily destroyed using lossy compression algorithm or a filtering process.
- Messages are hard to recover if image is subjected to attack such as translation and rotation.
- Robustness is low.

### III. PROPOSED ALGORITHMS

In this section we try to analyze few algorithms first we will look at the preprocessing.

*A. Preprocessing*

An image is selected as the cover image and the secret text is converted to binary and embedded in the image using embedding algorithm. Then the stego image is sent in the insecure transmission medium. Following figure represents the overall process of image steganography:

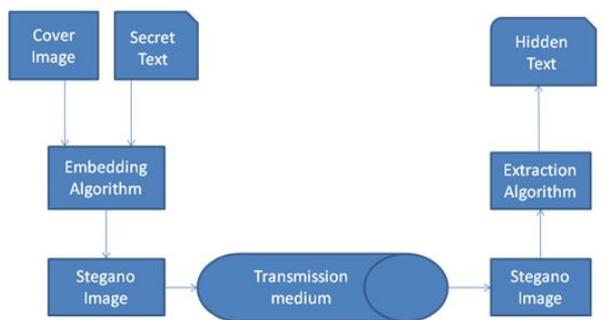

*B. Weighted matching technique:*

Image contains number of pixels each of which contains 3 bytes(Red, Green and Blue). The higher 3 bits of the lower nibble of each byte is interpreted to find the highest occurrence bit. This weighted bit is compared against a bit of the secret message under consideration. If the match is found, true is inserted in the LSB of the byte; else false is inserted.

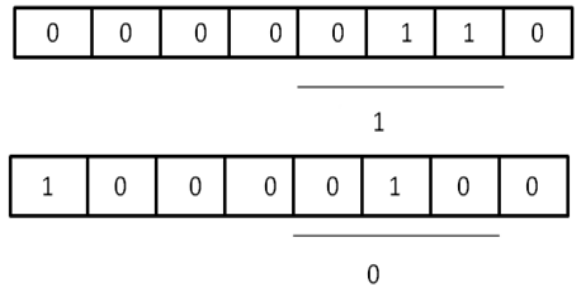

*C. Hiding Technique:*

The process starts by taking cover image and the secret text as input. They are converted into binary values and fed as input to the embedding algorithm. Next step is to check the size of the image and the secret text to confirm that the size of the secret text is not more than that of cover image. The following are the steps to be carried out in the proposed embedding algorithm.

Step 1: Start

Step 2: Select a RGB cover image and the secret text to be embedded.

Step 3: Check the size of the cover image and the hidden text.

Step 4: Encode the cover image and hidden text into binary.

Step 5: Choose a pixel and divide it into 3 bytes.

Step 6: Examine higher 3 bits of the lower nibble of a byte and find the highest occurrence bit.

Step 7: Compare the weighted bit against the bit under consideration of the secret text.

Step 8: If the bits are matched then insert 1 at LSB of the byte; else insert 0.

Step 9: Consider the next byte of the pixel and the next bit of the secret text. If the secret bits are embedded go to Step 9; Else go to Step 5.

Step 10: Stop.

Once the stego image is received at the receiver end, the secret message can be extracted by applying the reverse of the embedding procedure. The following are the steps to be followed in the extraction procedure.

Step 1: Start

Step 2: Select a RGB cover image and the secret text to be embedded.

Step 3: Encode the cover image and hidden text into binary.

Step 4: Check for highest occurrence bit in lower nibble excluding LSB bit.

Step 5: Check the value of LSB to get number n of value either 0 or 1.

Step 6: if n==0 then secret bit is negation of highest occurrence bit.

Step 7: Else if n==1 then secret bit is same as highest occurrence bit.

Step 8: Continue until all the secret bits are extracted.

Step 9: Stop.

The length of the secret message can be embedded in the buffer i.e. a separate pixels which are meant to contain system defined information. In the receiver side, at first, this information can be extracted there by knowing the length of the secret message to be extracted.

The following figure shows the embedding procedure:

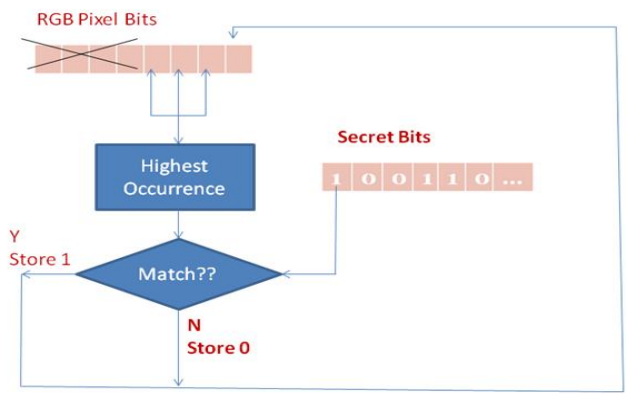

The extraction mechanism works opposite of the embedding procedure. It includes finding the weighted bit and the LSB of the byte; comparing them to get the actual message bit.

umber in which the 3 bits of the actual message lies.

## IV EXPREIMENTAL RESULTS AND DISCUSSION

For this algorithm, standard RGB color image 'Lenna' is used to measure the performance. The text message used to embed in the image is "Steganography is called covered writing".

The Figure 1 shows the original lenna image, Figure 2 shows stego image from the algorithm

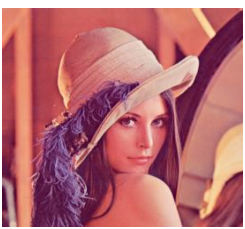    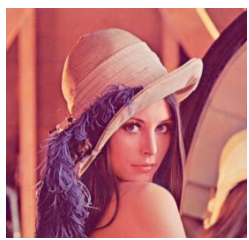

Figure 1                        Figure 2

The following table shows the PSNR value of the stego image by using the algorithm and traditional LSB insertion. Although distortion is not visible to naked eye, PSNR value indicates the quality of the stego image.

| Cover Image | Algorithm used | PSNR(dB) |
|---|---|---|
| Lenna.png | Proposed algorithm | 79.04 |
| Lenna.png | Simple LSB insertion | 78.65 |

. V. CONCLUSION

The proposed algorithm is better than the usual LSB insertion in the sense that we are not directly embedding the secret bit to the LSB rather we indicate whether the weighted bit matched secret bit or not. Also, the experimental results show that the proposed algorithm has better PSNR value than the usual LSB insertion technique.